# Study on Computational Thinking as Problem-solving Skill
## ~Comparison Based on Students' Mindset in Engineering and Social Science~

Andik Asmara
D10743015@yuntech.edu.tw
National Yunlin University of Science and Technology

Abstract
One of the capabilities which 21st-century skill compulsory a person is critical thinking and problem-solving skill that becomes top positions rank. Focus on problem-solving skills can be taught to a child, especially begun in elementary school refer to prior research focus on K-12. Computational thinking was one problem-solving skill that popular to implemented and studied in the current decade. This study was conducted to explore students' capability to be able solving of the problem based on the possibility use the computational thinking way. Participants in this study came from six international students that study in Taiwan and from two deferent sciences disciplines, engineering, and social science. A qualitative method was used to analyze data interview, took example cases from the global issue that is "Climate Change". The result founded that survive in a new environment was become evidence of their implementation of problem-solving skills. Problem-solving mindset both students of engineering and social science had discrepancy, those are how to use precise structure in the algorithm.

**INTRODUCTION**

The rapid development of technologies especially computer science bringing human mindset engagement, included in thinking skill (Lye & Koh, 2014; Moore et al., 2020; Papert, 1980; Resnick, 2007), creativity (Kim & Kim, 2016; Sadiku et al., 2019), and behaviors (NRC, 2011; Wing, 2006). Computer science was born in long-time ago; however, Wing (2006) has the renewed idea, boldly proclaimed the importance of a new skill-set and thinking ways that did not appear previously (Kolodziej, 2017). Further, a researcher in current decade raises the argument that Computational Thinking (CT) be able to be a fundamental skill for human, and breakout from computer scientists particularly (Wing, 2006; Kolodziej, 2017; Weese, 2013; Khine, 2018; Mouza et al., 2020). Computational thinking in daily life activities could be applying as problem-solving skills and can be started to teach children (Khine, 2018; Barr & Stephenson, 2011; X. Tang et al., 2020). Therefore, many researchers with various disciplines have innovation on how to implement CT and how to teach in various science fields.

The implementation of CT in the research field was mostly happening in education, for instance in the previous research has summarizing how to enhance CT abilities, solve problem use CT, teach CT using programing/coding, etc., (K. Y. Tang et al., 2020). Education is one of a field that pays attention to CT abilities, trying to enhance problem-solving skills starting from elementary school. Khine (2018) in their book explains that 37% of elementary students as populations in previous research (before 2018), then followed 28% undergraduate, 11% middle school and high school, and other grades. It means that CT implementation is almost available to each education grade and high possibility to merge with each subject to solve the complex problems. The reason teaches and trains CT in the education field that after graduation or accomplished the educations, they can implement in real-life or solve the problem in daily life.

As socialize human we faced a lot of problems in daily life came from internal and external factors (Zhang, 1991). To solve the problems, each person takes various solutions way base on their insight, experience, and environment. However, problem-solving skills can be learned from training, practice, and habituate innovative thinking. Computational Thinking as problem-solving skills could be taught to all ages (Khine, 2018). The importance is how to recognize the problem and then break down to several CT steps. CT it's a self-has various step to solve the problem based on previous researches. However, start in a couple of years ago be a focus on four-step that embed step every research, that is decomposition, pattern recognition, abstraction, and algorithm (Denning, 2009; Barr & Stephenson, 2011).

A computer as a tool is a powerful and useful equipment for processing and solving complex situations. Whereas as a thinking way is a thinking technic that uses concepts fundamental to computer sciences (Wing, 2006). Through several major steps and necessary the standard procedure in the solve problems. The complexities of the problem can be asunder to recognize the effectiveness of solving away. In addition, the superiority of this way is could be applying to a lot of field, issue, and multidiscipline sciences, in other word is borderless applying. Moreover, it's as well as could solve the problem that happening in daily life (Weese, 2013; Rich & Hodges, 2017; Bati et al., 2018) or the global issue in human life (Hunt & Riley, 2014).

Focus on computational thinking as problem-solving skills, and this research attracted to identify thinking skills or mindset a person. Based on global issues that everyone knows and trying to participate in a solver. Due to that, this study has purpose explore student's capability to able solving of the problem based on the possible use of computational thinking way as international students that know the issue. In addition, this study depth analysis and comparable engineering and social science student's mindset. In order to lead in reach of major purpose, this study raising the research questions that are:

- How does problem-solving ability to implement behavior activities as international students?
- What are the capabilities that be discrepancies in the possibility of computational thinking procedure implementation between students of engineering and social science?

## Literature review

### *Definition of "Problem-Solving" in Daily Life*

Discussion problem-solving context needed to know about meaning a problem first. Definition a problem based on the book by Jonassen (2004) described an unknown entity in some context, in other word has discrepancy between the goal and current state. Took from another journal had explained the problem is the treatment of activities to enforce the normal situation (Gibbs, 1965). Based on that, problem has means unnormal situation that giving uncomfortable situation (treatment) to human, and has effort to exit or jump-out from this situation called as problem-solving.

Problem-solving is a common part of our everyday experience (daily life) and founded everywhere for instance schools, universities, and socialization (Jonassen, 2004). This term needed predominant intellectual skill required humans in nearly vary settings. Proper (2012) in their article explain problem-solving as part of life, we cannot avoid mistake, and from the mistake, we can learn from them. Another reference describes problem-solving as ability skills learning, planning and concept development, and reasoning process (NATO Advanced Research Workshop, 1992). Whereas in daily life we face a problem and autonomous to learn from the problem then finally be able to solve it, called problem-solving skills.

### *Computational Thinking as Problem Solving Skill*

Learn problem-solving skills can be reached from many aspect fields, such as study cases, prior learning, and computer science particularly used computational thinking term. Through computer scientists be able to promote understanding of how to deliver computational process or structure ways to address problems in various fields (Barr & Stephenson, 2011). Learning computer science not just learn about the knowledge, were also achieve some problem-solving skills (Laski-smith et al., 2018). Parallel with the development of technology, computational processes are adopted into solving a problem in computer science self, engineering, social science, and daily life. Previous research (Park & Green, 2019) describes that Computational Thinking (CT) is a problem-solving skill that progressively more sophisticated as a person growth. Who learn CT they getting abilities included communication and collaboration skills, motivation, complex problem-solving skills, abstraction, and transfer (Rich & Hodges, 2017).

Wing (2006) explained that CT as technic to solving a problem, designing system plans, and to understand human behavior, based on the computer science fundamental concepts. The inside CT has a problem-solving skill set including the ability to: analyze the problem; reduce the problem complexity; develop an algorithm or plant for a solution; and verify the goal reached (Arfé et al., 2020; Román-González et al., 2017). Many previous studies provide describing CT components with various contents. Through researcher insight has experience in computer science to solve a problem involves four components or stages (BBC Bitesize, 2017), based on computer work principles. Four components (Rich & Hodges, 2017; BBC Bitesize, 2017) those included: Decomposition (breakdown problem to smaller parts (Arfé et al., 2020)); Pattern Recognition (find out the similarities among and within small part/problems); Abstraction (Focusing on important information only and ignoring irrelevant detail); and Algorithm (developing step-by-step /has rule solution to the problem) (Khine, 2018).

### *Possibilities of CT to Implement on Daily Life as Problem-Solving*

Based on computer science development has a role in all disciplines and daily life unexceptionally (Weese, 2013). One of the science disciplines in computer science that currently, rapid adoption as problem-solving skills is Computational Thinking. Begin 2006 computational thinking raising to be human attracted to implement in daily life, due to what Wing (2006) mentions that this field becomes a fundamental skill for everyone. In addition, she emphasizes this field also has ability to recognize human behavior by drawing on concepts fundamental to computer science. Human behavior included doing activities, communication, and thinking ways in daily life. Particularly, human thinking possible to identify the characteristic using computer ways principles. Depth considers in thought humans had a result raise new term that called computational thinking.

In the subchapter before, had explained computational thinking consist of four stages, which is Decomposition, Pattern Recognition, Abstraction, and Algorithm. Each stage can be called as represent human habit during does activities, especially in face challenges or a problem. This may unconsciously make the human self because autonomous likely. For instance, based on the book written by Hunt and Riley (2014) explain that in daily life in truth, each person following the algorithms process. Each human activity had does step-by-step process during reach the goal. Human autonomous learning in daily life as like as how a computer can solve a problem. It’s become a basis on computer solving principle ways that could be implemented as a fundamental skill in human daily life activities, not only on computer sciences (Wing, 2006).

## METHODOLOGY

The qualitative approach was used in this study to explore the student mindset in solving a problem. Besides catching the student mindset regarding solve the problem, this study as well as to recall the solution of problem experiences that provide by the qualitative research approach. Another else benefit from this approach is to have the ability to assess qualitative of thing using images, word, and description that described by the participant (Berg & Lune, 2017).

### *Research Procedure*

This study was used four steps procedure in during conduct the research. Step one is collect the raw data using interviews, second is transcription the data and categorize two groups (engineering and social science), third is exploration and analysis of the data using grounded theory, and the last fourth is to rise the conclusion based on finding and gaining the major theory.

Figure 1
*Research procedure*

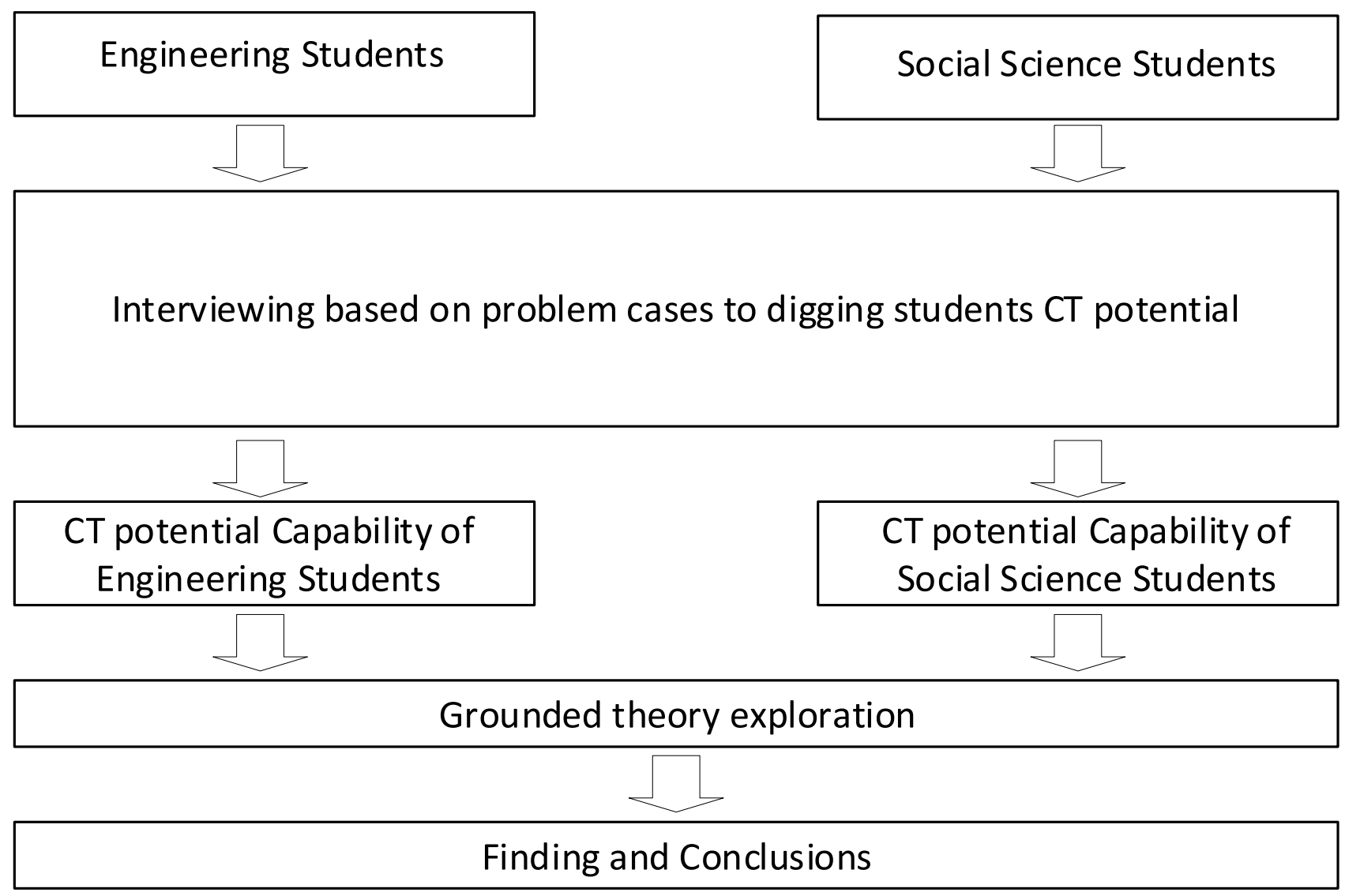

### *Samples*

This study was taking five participants that had status as an international student in one of the universities in Taiwan. The students came from three countries, which is Indonesia, Bangladesh, and Uganda. In order to select the participants was used cluster random sampling, divide two groups that are engineering samples and social science samples.

### *Data Collection*

This study was collected data use interviews, and before collecting the data also in order to align with the goal of this study, then interview guidelines are constructed. The interview guideline in this study can be a breakdown as follow:

- Prepare all things that needed in an interview before conduct interview
- Read, learn, and make clearer that related to major issue become interview topic, consist of:
    - Experience about solving the problem in daily life as students
    - Explain issue (climate change) that lead to thinking ability exploration
    - Raise a question about the major issue (climate change) included; do you know, causes, effect/impact, solution, structure to solve a problem
    - If have an unclear answer, then repeated the question and be more clear
- Foster willingness interviewee with make appointment and covering latter

This study took climate change as a case that interviewees supposed to try giving solutions based on their insight. The reason has chosen climate change it was used in previous research (Park & Green, 2019) to digging CT abilities.

Figure 2
*Data collection method*

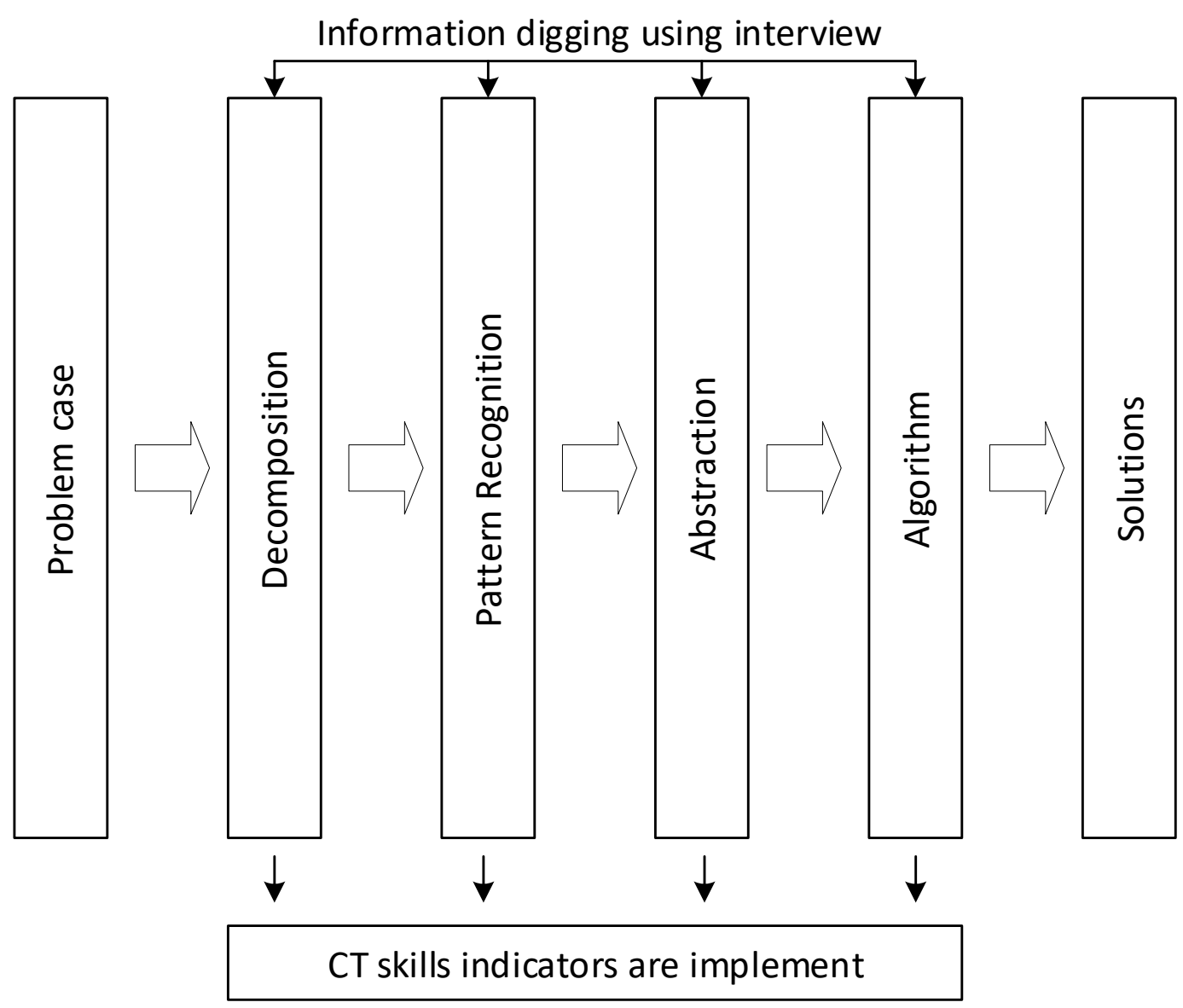

***Data Analysis***

The grounded theory analysis was used to looking for interviewees to answer equality to lead in a major purpose. This analysis has the possibility of a researcher to construct a conclusion based on the main majority answer.

**FINDING**

The five persons as respondents in this study were Doctoral student that had experience as an international scholar. The ages of respondents were various among 30 to 45 years old. The first stage was ensuring the interviewee knew about the issue of climate-changing and have a willingness to be interviewed. The impression of the interviewee is interested and fill-free to explain what they know. In order to simplify in identifying and exploring interviewee answers, in this study use code participant (P) to represent the interviewee.

***The Data Interpret of Daily Life Problem as Students***

The first stage in this study identifying and exploring the participant's experience regarding their problem experiences. The problem focuses on daily life problems faced in student activities. Each participant had an explanation in one or two cases that they draw to represent daily life problems. Most of a participant as new international students raises a problem in a daily activity outside the campus. The interpretation of the source the data could be written down as bellow;

*a. Environment changing*

Environment changing has means as students that different places on their study and place in their life before. For instance, P1 has a nation is Uganda, many things (included language, food, and culture) in daily activities and environment are differences. This happening in another participant, the cause comes from outside of Taiwan. In the detail of nations each participant is; P1 is Uganda, P2 is Bangladesh, and P3-P5 is Indonesia.

Language becomes a major problem that students faced. Chinese language was used in daily activities communications, and become challenging to new students that before didn't have study in this language. Several students interpret a problem as a challenge, its be motivating student to reach the goal of the challenge. Like as P1 answer that,

> *"Of course, normally as student we have challenge environment changing. Because I change from my country Uganda to Taiwan. So, it is totally different environment. New setting (every think is new) new food style, new people, new language, so is totally challenging …"*

In addition, almost all participant has this kind of challenge regarding language, foods, and cultures. In another viewpoint of the source of the problem, an international student that challenging from outside themselves. Due to moving from countries to stay, different flavor foods, and daily communication language.

After student's explanation their problem, then this study was collecting the data regarding how to solving? As a human have ways to solve the problem in daily life. Every person has the structure based on their abilities and capability to solving, further this study called mindset. Student mindset to solve the problem especially in this case of environment change that student prepares their self. Preparation starting from before coming to Taiwan and currently arrive in Taiwan. Student preparation includes mentally, organize themselves, and an open mindset to learn new things. Like as P1 and P5 explanation in their answers as,

> *P1: "I had to organize myself. I had to understand …"*
>
> *P5: "We should well prepare for a mentally, adjust to adapt in Taiwan".*

*b. New insight in major study*

Further, new insight into the major study of a student becomes a challenge for him. Learning new things supposed new students to equipped previous insight changes after determining the major study. This necessary to students more capable and become experts as doctoral students in one focus area. Like a P3 explanation in their facing a challenge,

> *"… problem that I face has two type, that is academic and non-academic. In academic regarding with major study of our research…"*

In order to solve this problem as a new doctoral student have self-management, and they pursue a lack of insight discrepancies. Learning new things are needed in a study on the doctoral degree to support their research. This was explained P3 as bellow,

> *"… my knowledge background still deficient, therefore I do matriculation myself to pursue of the lack regarding insight that needed."*

***The Data Interpret in CT Stages***

Before jump into computational thinking ability interpretation, quiet importance this study digs out insightful of the student, especially in raise the issue. In the beginning, this study was digging out of insight into climate change for each student. The result was all the students had insightful about this issue. Seemly very confident the respondent explains the means of climate change.

The challenge in this study is interpreted data from the student mindset to the computational thinking abilities of each stage. Computational thinking has four steps, that are decomposition, Pattern Recognition, Abstraction, and Algorithm. The interpretation of CT stages to solving a problem this study based on the researcher's insight and the literature that supports.

1. *Decomposition*

The first stage is decomposition, this stage aims to identify an approach to student mindset. Focus on the ability to recognize the major problem, then breakdown to several small problems that relevant. The result of interviewee data transcription and then interpret to decomposition write down in the table below. In order to the guidance into decomposition stage has provided decomposition categorize in some sub-theme.

Decomposition 1: Causes of Climate Changing;

*P1: "… there are many problems or there are many causes that could lead to climate change, .... One of the challenges we have as the media calls climate change is deforestation. .... Another aspect that can cause to climate change is the pollution."*

*P2: "… carbon dioxide is increasing, oxygen is decreasing (PR1). That is the main cause for this climate change."*

*P3: "increase in pollution, such as CO2 increment"*

*P4: "… industries sector, there a lot of has pollutions, and one of sector that contribute to pollutions."*

*P5: "The sector of industri was used hazardous material, and human behavior as well contribute to climate changing."*

Decomposition 2: how has responsibility;

*P1: "... I think we all are responsible because we are all members of the ecosystem. ... the government has to be there."*

*P2: "... the developed country because the developed country there have many, many industries …"*

*P3: "responsibility all of us, which one is government as regulator."*

*P5: "Human self, …. and government as representative responsibility."*

Decomposition 3: Have solutions to solving;

*P1: "... first, we need to sensitize the masses about their collective responsibility, about environment. We need to ensure ecofriendly activities ..."*

*P2: "... I will request to every people they must plant some trees, ... Another way, I will make some rules for the developed country. And they must follow the rules to reduce the carbon dioxide."*

*P3: "… as government involving every sector (industries/civilization) get in contribution to solving this problem."*

*P4: "… definitely involving around the word government to in charge of real actions."*

*P5: "… establishing a policy, such as limitation of private transportation to decreasing pollution …"*

In addition, not limited just on three sub-themes above to decompose the general problem to a small problem. It is depending on the solving ways every person, each person have themselves ways based on their insight.

2. *Pattern Recognition*

Pattern recognition is a stage that a person recognizes the special or unique characteristic of each category of decomposition. Focus on this issue climate change has several characteristics that special and unique, and this characteristic become part of the puzzle to solve the problem. Based on interviewee result, the categorize as pattern recognition as bellow;

Pattern Recognition 1: Causes of Climate Changing;

*P1: "Pollution kind be water pollution, air pollution …."*

*P2: "… carbon dioxide is increasing, oxygen is decreasing."*

*P3: "… increase the CO2."*

*P4: ''… air pollution from industries."*

*P5: "… motor vehicle fumes increase ..."*

Pattern Recognition 2: how has responsibility;

*P1: "… government is one of the stakeholders of the ecosystem. .... We need to be responsible of our activities. ...."*

*P2: "… the develop countries pursue largest industrial development."*

*P3: "… the government has responsibilities to control and establishing policy."*
*P5: "… the government have power as representative position."*

Pattern Recognition 3: Have solutions to solving;
*P1: "... manage the wastes."*
*P2: "… plant trees."*
*P3: "… energy saving."*
*P4: "… largest reduction."*
*P5: "… motor vehicle restrictions."*

3. *Abstraction*

Abstraction in the context of problem-solving where each small problem that previously does breakdown had a solution. The solution was used person to express or interpret what they do to solving a problem. In the interview result, the abstraction could be a sentence that has means functions or the ability to offer to solve a problem. As bellow result in might categorize as an abstraction;

Abstraction 1: Causes of Climate Changing;
P1, P3, P4: "*Pollution*"

Abstraction 2: how has responsibility;
P3: "*As a regulato*r"

Abstraction 3: Have solutions to solving;
P1: "*Eco friendly activities, public awareness*"
P2: "*Awareness*"
P3: "*Energy saving*"
P4: "*Reducing*"
P5: "*Conventions*"

4. *Algorithm*

The solutions that offer each small problem breakdown need to be sort based on which one is solved first to and the end. Here a problem solver supposed to use a strong logic to determine the phased stage actions and sequential work. Khine (2018) named this stage as algorithmic thinking, that has meant devising a step-by-step solution to a problem and differs from coding. This stage is hard seemly to identify algorithms in each answer that collected. As bellow was algorithm interprets from interviewee answer;

P2: "*Due to pollution → government as regulator establish policies → planting tree and foster awareness*"
P3: "*Due to increase CO2 and decrease O2 → A country that is government → involved citizen to do energy saving and foster environment awareness*"
P4: "*Due to pollution → Civilization has participated → pollution reduction*"
P5: "*Due to pollution → government as regulator → motor vehicle restriction*"

Table 1.
*Coding participants answer into CT context*

| | | Engineering | | | Social Science | |
|---|---|---|---|---|---|---|
| CT components | Breakdown | P2 | P3 | P4 | P1 | P5 |
| Decomposition | D1 - What are the Causes<br>D2 - Who has responsibility<br>D3 - How are to solve | D1, D2, D3 | D1, D2, D3 | D1, D3 | D1, D2, D3 | D1, D2, D3 |
| Pattern Recognition | PR1 - know the causes<br>PR2 - know the responsibility<br>PR3 - know to solve | PR1, PR2, PR3 | PR1, PR2, PR3 | PR1, PR2 | PR1, PR2, PR3 | PR1, PR2, PR3 |
| Abstraction | A1 - term of causes<br>A2 - term in responsibility<br>A3 - term of solving | A3 | A1, A2, A3 | A1, A3 | A1, A3 | A3 |
| Algorithm | Al - used structure from causes, responsibility, the solver | Al | Al | Al | Al | - |

## DISCUSSION

This study took climates changes as a problem or a case that offer to the respondent. The reason is this issue a global issue (Lal, 2004) that everyone knows, moreover as international students. Each person has a solution based on their insight related to the global situation, particularly increase the degree of temperature (Nordås & Gleditsch, 2007; IPCC, 2011) and increase the level of sea waters (Taylor et al., 2013). The temperature degree increases one of that due to pollution that happening in human daily life (Linden & Office, 2015; Jacob & Winner, 2009; IPCC. Painel Intergovernamental sobre Mudanças Climáticas, 2014). This theoretical previously becomes a guideline to judge and draw an interpretation of the interviewee's answer.

### *Implementation Problem Solving Ability into Daily Life*

Before discussion focus on the major topic to interpreted student ability in solving a problem that offered, this discusses first regarding student problem-solving ability to implement in daily activity as a student. Problem-solving became an ability that persons have, in order to face current daily life and complexity in the life. Problem-solving based on theoretical previously consists of two phases, that is problem representation and problem solution (Huang et al., 2019). Problem representation is how people recognize situations or conditions that face being a problem. People usually categorize a problem from easy to hard, and urgently to not urgently. Several people assume that easy problems unrecognized as a problem. Further, they just assume as resistance in daily life activities, however, can quickly solving. This happens as well on respondent fifth that they mention as follow:

*"I feel no any problems, just adapt in Taiwan environment and conditions."*

It means they have high confidence in the face of the current situation and well prepare a study in Taiwan. Several mostly participants in this study have problems that they need to go out of this situation and solve this uncomfortable feeling. The common situation that students faced in a new environment and then assume as a problem, that is food, language, and cultures. This can predict each new international student in the new environment and situation has a similar problem.

In order to solve those kinds of problems students took action to adapt to a new environment. Based on the result of the interview process, find out that learn from another person or learn from the environment takes the position in importance place. The keyword as problem-solving function just use one word that is “learning”. According Jonassen (2004) in the book entitled “learning to solve the problem” explains that human activity which called learning is driven by problem that needs to solve. Still, this book introduces the function of learning, that is oriented by an intention to resolve the dissonance, satisfy the curiosity, answer the question, and figure out the system. Learning is a powerful function to effective implementation as a problem-solving way in behavior daily life activities.

***Engagement Computational Thinking into Problem Solving Mindset***

Talking about major themes in this study, which interprets a person's mindset based on the side view and insight of a researcher, particularly in the major topic discussion. This research depth analysis on a person's mindset to solving a problem and how to solve ways would be engaged with computational thinking stage structures. In order to clearly understand CT concept to solving e problem, currently, discussion divide consists of four-part as bellow.

Figure 3.
*CT stages illustration*

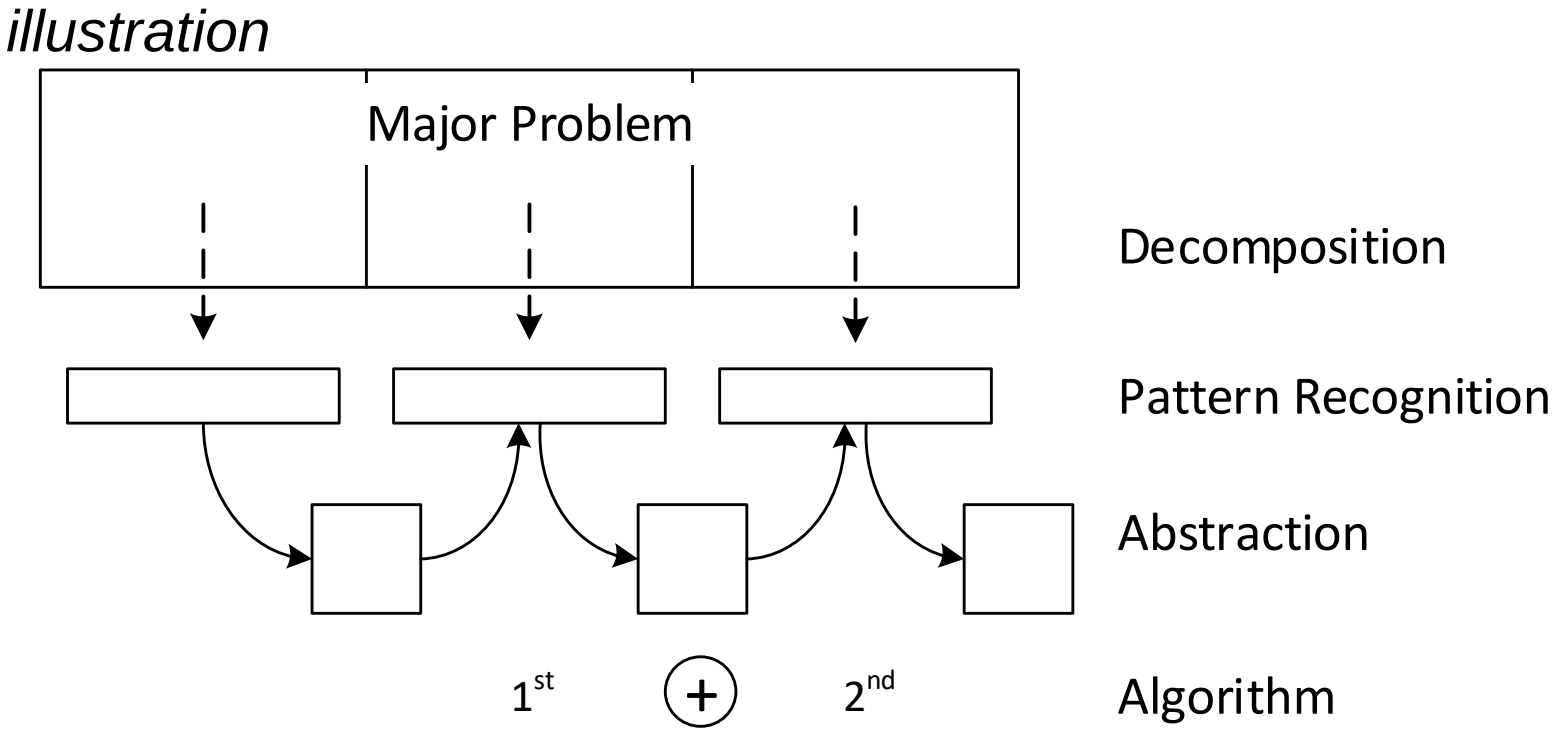

Focus on the case in this study that is climate change has complexities and come from several causes. Climate changing as a problem comes from accumulation causes that unsolved. Whereas can divide or separate into several small problems. This dividing process called decomposition (stage-1) process from a big problem to several small problems. Park & Green (2019) describe that decomposition is breaking down the task (problem) into smaller or manageable parts. Turn back to interpret the interviewee's answer, which includes process decomposition are a student has ability to mentioning and explaining “causes”, “has a responsibility”, and “how to solve”. Those three keywords represent the decomposition process (breakdown in three parts) from major problem climate changing.

A further stage is identified as a unique and linkable characteristic in each breakdown problem. In computational thinking, a science called pattern recognition (stage-2) and used to find the solutions. Kale (2018) explains that this stage is to identify patterns among smaller parts. Another reference mentions the ability to identify and match similarities in each smaller part (Kynigos & Grizioti, 2020). From the interview result, what kinds of unique characteristics to be one of the similarities each smaller part? The result of the interviewee answer has unique or characteristic in one word, that is “pollution”. This characteristic as well becomes a fact that causes climate changing (Linden & Office, 2015; Jacob & Winner, 2009). The reasons “pollution” become the pattern for all part are; (1) causes pollution: increase CO2, Industrial air pollution, motor vehicle pollution; (2) have role to control pollution: government, a country; and (3) reduce pollution: plant tree, manage waste, and motor vehicle restriction.

The next stage is finding out the solution on a small part. In addition, we just use the solution (function) to give a solution in a small problem and called Abstraction. From previous research, abstraction has the meaning of reducing complexity to define my idea (Park & Green, 2019). It means the complexity has the potential to be a problem that will be solved or answer use this stage. From the result of the interview, each small part has a word that becomes the abstraction, has meaningful, and simplifies from complexity a problem. There are words become abstraction; “pollution”, “as a regulator”, and “awareness”.

The “pollution” has means complexity that happening because of a combination of high emissions and unfavorable weather (Jacob & Winner, 2009), un-managed waste (Marshall et al., 2013), and the effect of industrialization as well as increase motor vehicle. Whereas “as a regulator” has means they have the power and authority to establish policies regarding climate change addressing. Such as waste management, motor vehicle restriction, and renewable energy usability. In addition, “awareness” has meant their capacity to adapt to climate change risk (Marshall et al., 2013), preventive action, and addressing.

The last stage in the Computational Thinking process is using the solver function in the unity step and bringing the solution to whole a big problem. This process step called Algorithm has meant a set of instructions for humans or machines to follow to solve a problem (Moore et al., 2020). A set of instructions refers to a structure of solving functions that come from the previous step, which is an abstraction. This stage had result came from an analysis of participants answer and draw the algorithm figure based on researcher insightful. Based on the result of the interviewee, the majority participant has ability to construct the structure of solving a major problem. If generalize the answer algorithm that participant used could be drawn as follows;

Figure 4.

*Algorithm stage came from abstract combination*

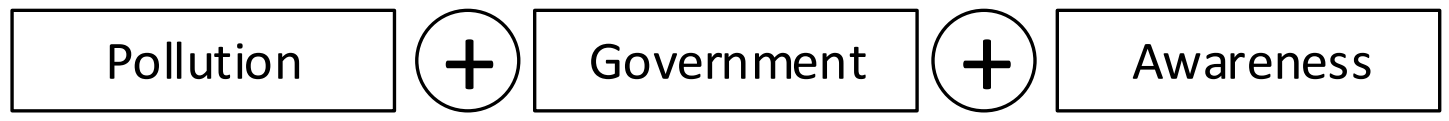

In order to solve a climate change problem, among three-part of solver functions are implemented together and each other have mutual support. The first algorithm step is what kind of pollution causes, then the second step is which one government that has a responsibility, and the third step is how to foster awareness in the environment to decrees pollution. Those three-kind function as solution each problem breakdowns to small parts. Whereas, the majority respondent can answer as solving a problem seemly refer to figure 4. The combination and structure of the word “pollution”, “government”, and “awareness” become algorithms as problem-solving for climate changing.

***Draw Student Problem Solving Ability into a Figure CT Context***

In order to answer the purpose of this study, the researcher tries to draw of comparison of the solution answer between engineering and social science students. Therefore, the categorize student answer into table 1 based on the result of what this study got. Researchers try to interpret from sentence description to quantitative result to create figures as represent a comparison between both of science. The result of interpreting participants' answer to the CT context can see in figure 5.

Figure 5.

*Comparison CT skill in Engineering and Social Science Students*

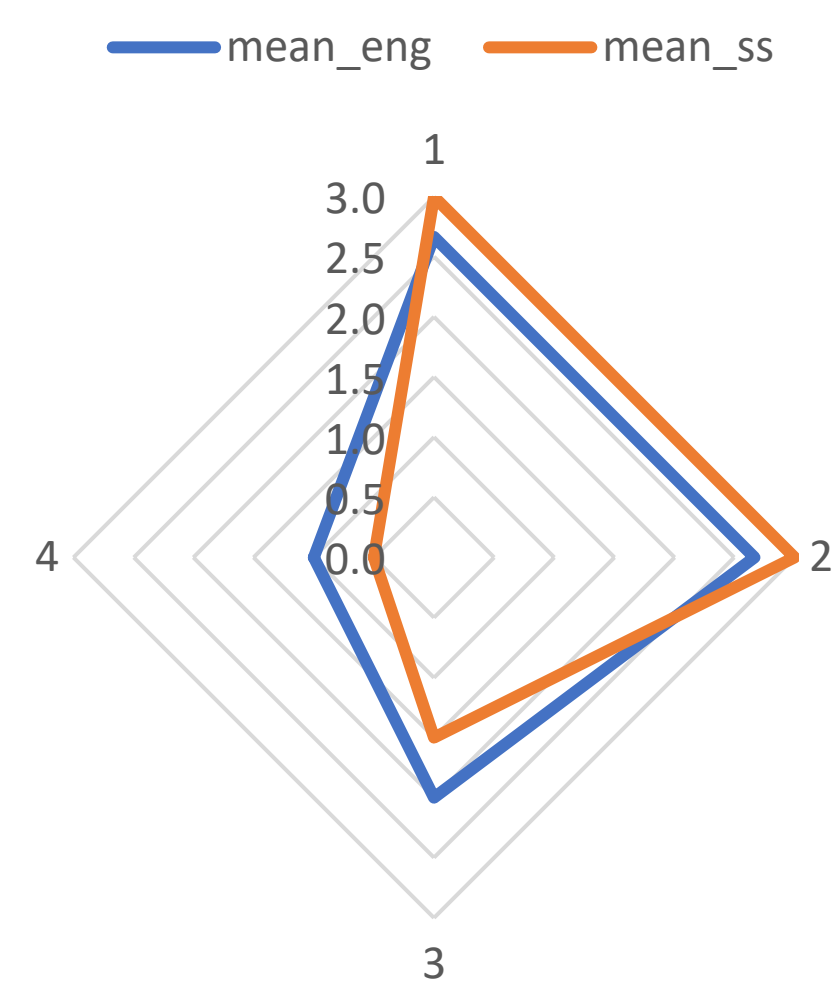

The “mean_eng” has meant the interpret data came from engineering international students, and “mean_ss” is the interpret data that came from social science international students. The result showed that in decomposition and pattern recognition social science students little higher in describing the problem. Whereas in the abstraction and algorithm engineering students higher in find out the solution and used solutions in the unity structure. It has meant that engineering student has the ability to use syllogisms to make good design decisions in their problem faced (Holvikivi, 2007). This happens due to engineering student had usually work and do something in their field used logical and algorithm.

## CONCLUSIONS

The international Doctoral students in the studies their learn about some problem and trying to solve that. In daily life, they have challenges that must face being adapt in new environments. This study showed that as Doctoral students, they know how to solve the problem. It's proven by evidence that they can survive in a new environment, culture, and challenge during a study in Taiwan. Several students in inconvenient situations just assume that as a challenge or not like a problem. Whereas other students, they assume as a problem, however still under-control and can solve it.

In the facing a problem (for instance: climate change) all student as participants, they can recognize the problem and decompose the problem to several small problems. Whereas to compare in thinking technic (focus on Computational Thinking) between engineering and social-science students had a discrepancy in how to solve the use of precise structures (algorithm). An engineering student has a better structure rather than social-science students on how to explain the problem, decomposition, pattern recognition, abstraction, and finally, use in the algorithm. It happens due to engineering students usually use logical thinking in their work and structure steps in to solve problems.